\documentclass[journal=jpccck,manuscript=article]{achemso}

\usepackage[utf8]{inputenc}
\usepackage[T1]{fontenc}
\usepackage{mathptmx}

\usepackage{graphicx}
\usepackage{dcolumn}
\usepackage{bm}
\usepackage{hyperref}
\usepackage{upgreek}
\usepackage{dcolumn}
\usepackage{bm}
\usepackage{amsmath}%
\usepackage{amssymb}%
\usepackage{xcolor}
\usepackage{nameref}

\author{J. Pedro de Souza}
\affiliation[Chemical]
{Department of Chemical Engineering, Massachusetts Institute of Technology, 25 Ames St, Cambridge, Massachusetts 02142 USA}
\author{Martin Z. Bazant}
\affiliation[Chemical]
{Department of Chemical Engineering, Massachusetts Institute of Technology, 25 Ames St, Cambridge, Massachusetts 02142 USA}
\alsoaffiliation[Math]
{Department of Mathematics, Massachusetts Institute of Technology, 182 Memorial Drive, Cambridge, Massachusetts 02142 USA}
\email{bazant@mit.edu}

\title {Continuum theory of electrostatic correlations at charged surfaces}

\begin{document}

\begin{abstract}
The standard model for diffuse charge phenomena in colloid science, electrokinetics and biology is the Poisson-Boltzmann mean-field theory, which breaks down for multivalent ions and large surface charge densities due to electrostatic correlations. In this paper, we formulate a predictive continuum theory of correlated electrolytes based on two extensions of the Bazant-Storey-Kornyshev (BSK) framework:  (i)  a physical boundary condition enforcing continuity of the Maxwell stress at a charged interface, which upholds the Contact Theorem for dilute primitive-model electrolytes, and (ii) scaling relationships for the correlation length, for a one-component plasma at a charged plane and around a cylinder, as well as a dilute z:1 electrolyte screening a planar surface. In these cases, the theory accurately reproduces Monte Carlo simulation results from weak to strong coupling, and extensions are possible for more complex models of electrolytes and ionic liquids.
\end{abstract}
\maketitle
\section{Introduction}
Electrostatic correlations can significantly affect the structure and thermodynamic properties of the electrical double layer~\cite{Levin2002,YuGrosberg2002}, resulting in qualitative differences from mean-field Poisson-Boltzmann (PB) theory, such as like-charge attraction \cite{pellenq1997electrostatic, misra2019theory} or over-screening of surface charge. Critical applications in biology, colloids, separations, or electrochemistry rely on or operate in the regime where correlation effects are critical.

Numerous models have been proposed to capture electrostatic correlations, typically with a complicated mathematical structure. Outwaithe derived a modified Poisson Boltzmann models to account for the fluctuation potential of a single ion interacting with a charged wall \cite{outhwaite1983improved}. The hypernetted chain approximation and mean spherical approximation closure to the Ornstein-Zernike equation involve solving integral equations involving the direct correlation functions of bulk charged spheres \cite{henderson1978some, henderson1979application, lozada1982application, kjellander1992double} for the equilibrium structure. Further work based on classical Density Functional Theory determines equilibrium properties based on the minimization of an integro-differential free energy functional.  Kierlik and Rosinberg implemented a model (termed the bulk fluid model \cite{voukadinova2018assessing}) which captures correlations based on an perturbative expansion of density with direct correlation functions as an input from the mean spherical approximation \cite{kierlik1991density}. Voudkavinova et al. analyzed the accuracy of two other related density functional theories (reference fluid density \cite{gillespie2002coupling}, functionalized mean spherical approximation \cite{roth2016shells}, and bulk fluid \cite{ kierlik1991density, rosenfeld1993free})  in comparison to Monte Carlo simulations, finding the reference fluid density approach  to be most accurate \cite{voukadinova2018assessing}. These theories implement the accurate Fundamental Measure Theory functional developed by Rosenfeld to describe excluded volume effects \cite{rosenfeld1989free, roth2010fundamental}. The additional electrostatic interactions beyond mean field are included in the excess free energy separately, without any modification to the mean-field electrostatic part of the free energy. 

While these approaches often produce accurate density profiles, they can be involved to implement to a broad range of applications, for different geometries or dynamic problems, especially compared to the classical PB theory. To our knowledge, they also have not yet been shown to recover the correlated behavior of the counterion-only limit for counterions of infinitesimal size. A simpler mathematical structure could help with the application and interpretation of electrostatic correlations to a wider class of problems in physics, including electrokinetics, colloidal interactions, and electrochemical transport/reactions.

\begin{figure}
\centering
\includegraphics[width=0.5\linewidth]{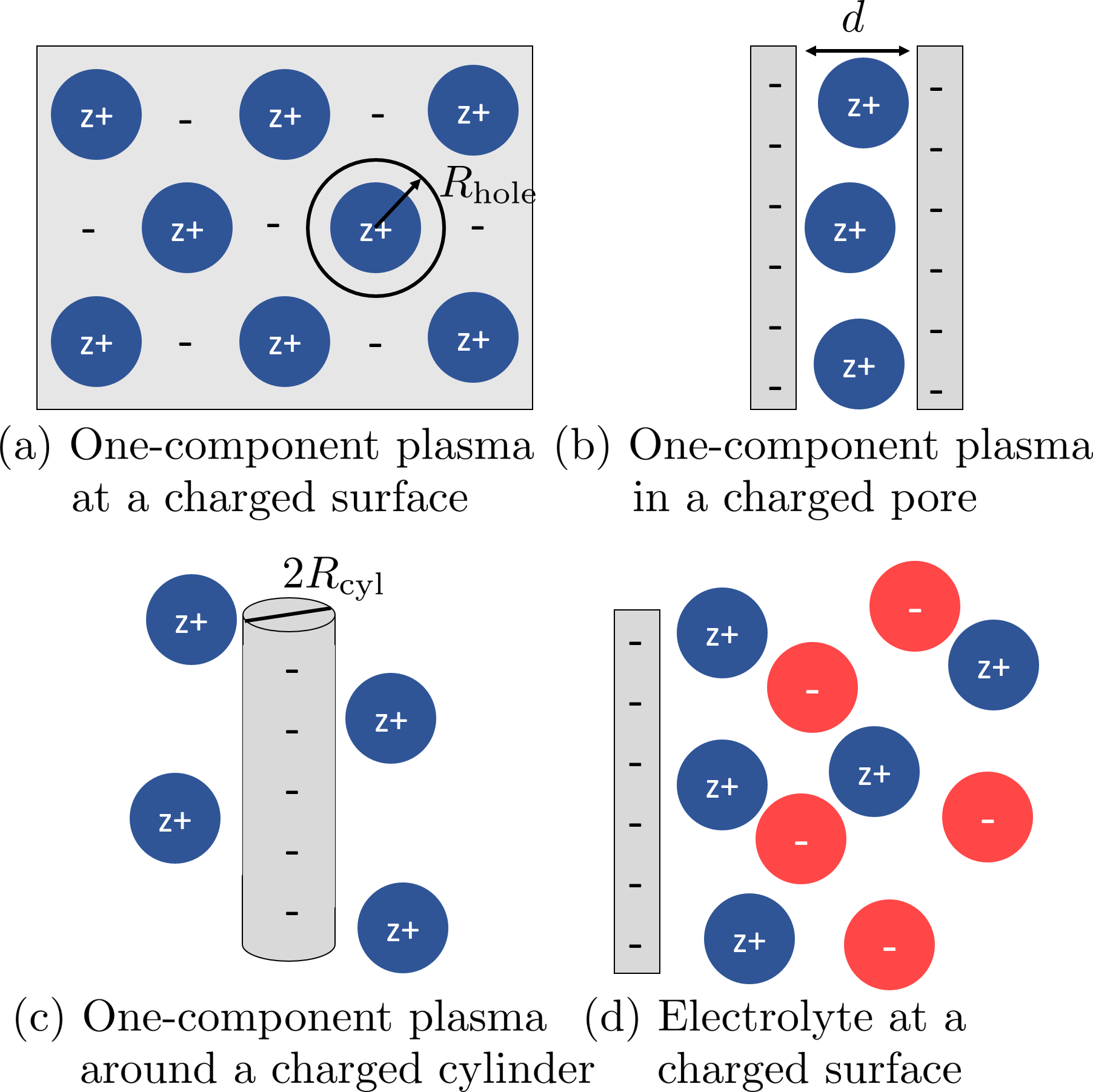}
\caption{The scenarios considered in the application of the BSK theory.}
\label{fig:fig1} 
\end{figure}
Bazant, Storey and Kornyshev (BSK) proposed a continuum framework to account for the nonlocal dielectric permittivity of ionic liquids resulting from ion-ion correlations~\cite{Bazant2011} with a simple mathematical structure, building on intermediate coupling approximations of Santangelo~\cite{santangelo2006computing} and Hatlo and Lue~\cite{Hatlo2010} for the one-component plasma. The model captures correlations based on expansions in terms of \textit{electric field}, rather than ion density, in the free energy functional which leads to a higher-order Poisson equation. In so doing, the electrostatic correlations are included self-consistently in the definition of the electrostatic potential whose gradient determines the electrostatic force on an ion in the diffuse layer. The BSK theory provides a simple framework to predict charge density oscillations and over-screening phenomena in a variety of electrokinetic , electrochemical, biophysical, and colloidal phenomena in electrolytes and ionic liquids. The equations require a similar level of complexity to solve compared to the PB theory, which allows them to be applied to a broad group of applications. The theory was used to describe electrosmotic \cite{storey2012effects} and electrophoretic mobility \cite{stout2014continuum} reversals in multivalent and concentrated electrolytes, as well as electroconvective instabilities in ionic liquids \cite{wang2017modeling}. It was also applied to the dynamics \cite{zhao2011diffuse, jiang2014dynamics, kondrat2015dynamics, alijo2015effects, alidoosti2018impact} and electrosorption \cite{lee2013electric, mceldrew2018theory, shalabi2019differential, xie2018nonlocal} at electrochemical interfaces for ionic liquids and concentrated solvent-in-salt electrolytes, including storage \cite{yang2019discrete} and transport \cite{jiang2016current} in nanoporous media. Electrostatic correlations have a profound effect on colloidal interactions\cite{misra2019theory, moon2015osmotic, santos2017effect}, where they can induce like-charge attraction in multivalent electrolytes, also predicted by the BSK theory. The activity and solvation energy of electrolytes at high concentration was also studied including the electrostatic correlation effect \cite{liu2015poisson2, schlumpberger2017simple,liu2018poisson,nakamura2018effects,liu2019generalized} as well as the extent of ion pairing in confinement\cite{huang2018confinement}. Finally, the electrostatic correlations given by the BSK theory were important in describing the conduction through biological ion channels \cite{liu2013numerical, liu2013correlated,liu2014poisson, Liu2014, liu2015numerical}. Despite the numerous applications, fundamental questions remain about the proper boundary conditions and correlation length required to complete the BSK theory.

Here, we show that the appropriate boundary constraint for the higher-order Poisson equation is based on an interfacial stress balance. With  corrected boundary conditions, the BSK theory becomes exact in the strong and weak coupling limits for the one-component plasma, and agrees with Monte Carlo (MC) simulations in intermediate coupling. We also suggest scaling relationships for the correlation length without steric constraints in one-component plasma and in multivalent electrolytes, for all the scenarios in Fig. \ref{fig:fig1}. We show how the correlation length can have a simple physical interpretation based on the correlation hole size of counterions at a charged surface. Although generalizations are possible, we restrict the analysis to a restricted primitive electrolyte with hard, spherical ions of equal size in a constant permittivity, $\epsilon$, medium and  smeared out surface charge density, $q_s$, and neglect all concentrated-solution effects, so as to isolate electrostatic correlations.
 
\section{Theory}
The BSK free energy functional is given by:
\begin{equation}\label{eq:eqFreeEnergy}
\begin{split}
G=&\int_{V} d\textbf{r}\, \left\{ g+\rho \phi -\frac{\epsilon}{2}\left[\left(\nabla\phi\right)^2 +\ell_c^2\left(\nabla^2\phi\right)^2+\dots\right] \right\}\\
&+ \oint_S d\textbf{r}_\textbf{s} \, q_s \phi.
\end{split}
\end{equation}
Here, $g=(H-TS)/V$ is the enthalpy and entropy density, $\rho$ is the charge density, and $\phi$ is the electrostatic potential. For simplicity, the free energy is truncated after the first correlation contribution, although higher order terms can be considered \cite{Hatlo2010}. While the original authors performed a gradient expansion to arrive at Eq. \ref{eq:eqFreeEnergy} \cite{Bazant2011, storey2012effects}, the mathematical procedure is equivalent to modifying the interaction potential between ions from $U_{\alpha\beta}(r)=z_\alpha z_\beta \ell_B\mid r\mid^{-1}$ to $U_{\alpha\beta}(r)=z_\alpha z_\beta \ell_B\mid r\mid^{-1}(1-e^{-\mid r\mid/\ell_c}) $ \cite{kondrat2015dynamics, santangelo2006computing}. The modified interaction potential is solved in the mean-field limit. Thus the BSK theory is a phenomenological correction to PB within the mean-field approximation, by subtracting out interactions with smeared out charges within a correlation length $\ell_c$, which should scale as the size of the correlation hole. The modified Poisson equation results by finding the extremal of the functional ($\frac{\delta G}{\delta \phi}=0$): 
\begin{equation} \label{eq:BSKeq}
\epsilon(\ell_c^2\nabla^2-1)\nabla^2\phi=\rho.
\end{equation}
Eq. \ref{eq:BSKeq} is a statement of Maxwell's equation, $\nabla\cdot \mathbf{D}=\rho$ where the displacement field is $\mathbf{D}=\hat\epsilon\mathbf{E}$ with a non-local permittivity operator $\hat{\epsilon}=\epsilon(\ell_c^2\nabla^2-1)$ applied on the electric field, $\mathbf{E}=-\nabla\phi$, in a medium of constant permittivity, $\epsilon$. PB theory is recovered when $\ell_c=0$. Note that the definition of electrostatic potential itself has changed by adding the higher order correlation terms, without violating Maxwell's equation. In other words, the potential that determines the energy of an ion in the double layer must satisfy $\nabla\cdot\left( \hat\epsilon\mathbf{E}\right)=\rho$ rather than $\nabla\cdot\left( \epsilon\mathbf{E}\right)=\rho$. In this way, the electrostatic correlation contribution to the electrostatic energy for an ion is included self consistently within the electrostatic framework, rather than being added as additional corrections in the excess chemical potential. An advantage of this approach is that the electrostatic force per unit charge of an ion is captured directly with $\mathbf{E}$, meaning that the diffuse potential $\phi$ here could be measured experimentally at an electrode (if also accounting for the potential drop in the Stern layer).

The charge density at equilibrium, $\rho=\sum_i z_iec_i$, will be determined by the constraint that the electrochemical potential for each ion is a constant at equilibrium. The electrochemical potential can be defined as the variation of the Gibbs free energy with respect to concentration \cite{bazant2013theory}, $\mu_i=\frac{\delta G}{\delta c_i}$, or 
\begin{equation}
    \mu_i=\mu_i^\theta+kT\ln(c_i)+z_ie\phi+\mu_i^{\mathrm{ex}}
\end{equation}
where the first term is a reference value, the second term is the ideal entropy contribution, the third term is the electrostatic potential contribution, and the fourth term is the excess electrochemical potential. 

The first open question in applying BSK theory is that of additional boundary conditions, beyond Maxwell's equation $\hat{n}\cdot \mathbf{D}=q_s$. Presumably, the boundary condition must take care of the unaccounted short-range part of $U_{\alpha\beta}$. In the original BSK formulation and all subsequent works, the boundary condition of $\hat{n}\cdot\nabla^3\phi=0$ was applied, with the justification that the correlation effects should disappear at the interface \cite{storey2012effects, kondrat2015dynamics,jiang2014dynamics, jiang2016current, huang2018confinement, Santos2009, stout2014continuum, moon2015osmotic, liu2013numerical, alijo2015effects, lee2013electric, alidoosti2018impact, mceldrew2018theory, gupta2018electrical}. The theory provided reasonable agreement to simulation and experimental results for ionic liquids and multivalent electrolytes. However, the boundary conditions have not yet been proved or validated systematically.  

The second open question is the choice of correlation length, which has been arbitrarily set to the Bjerrum length for electrolytes \cite{Bazant2011,stout2014continuum}, and the ion diameter for ionic liquids \cite{Bazant2011}. The theory is ultimately very sensitive to the choice of boundary conditions and correlation length. Here, we analyze the boundary condition in terms of a stress balance at the interface and then validate $\ell_c$ by comparison to MC simulations.

\subsection{Interfacial balance}
Applying the Gibbs-Duhem equation at constant temperature to the electrolyte and screened surface charges, following \cite{Bazant2009towards}, and neglecting the external electrostatic work done on the system, gives $dP=\sum_i c_i d\mu_i$. Taking the gradient in three-dimensional space and applying the definition of the electrochemical potential:
\begin{equation}\label{eq:eqGibbsDuhem}
    -\mathbf{f}=\nabla P= k T \sum_i \nabla c_i + \rho \nabla\phi+\sum_i c_i \nabla\mu_i^{\mathrm{ex}}, 
\end{equation}
where $\mathbf{f}$ is the total thermodynamic force. The first and third terms on the RHS of Eq. \ref{eq:eqGibbsDuhem} correspond to the gradient of osmotic pressure, $\nabla \Pi$.  For an ideal solution, $\mu_i^{\mathrm{ex}}$=0. The gradient of the defined thermodynamic pressure is equivalent to the divergence of the total stress tensor of the electrolyte system, $\mathbf{f}=\nabla\cdot {\mathbf{\uptau}}$. The total stress tensor is composed of an osmotic pressure component, $\Pi$ and a Maxwell stress tensor, $\mathbf{\uptau}_e$, such that $\mathbf{\uptau} = -\Pi \, \mathrm{I} +\mathbf{\uptau}_e$. The component of interest in this analysis, $\mathbf{\uptau}_e$, can be defined by: 
\begin{equation}
    \nabla \cdot \mathbf{\uptau}_e = \rho\mathbf{E}= \nabla \cdot(\hat\epsilon\mathbf{E})\mathbf{E}
\end{equation}
in a constant $\epsilon$ medium. Plugging in for the charge density using the BSK Eq. \ref{eq:BSKeq} and performing integration by parts, one arrives at an expression for the Maxwell stress tensor for a fluid with a non-local permittivity,
\begin{equation} \label{eq:MaxStress}
\begin{split}
    \mathbf{\uptau}_e=& \epsilon \mathbf{E E}-\frac{1}{2}\epsilon\mathbf{E}^2\,\mathbf{I} 
     +\epsilon {\ell_c}^2\Big[\left(\mathbf{E}\cdot \nabla^2 \mathbf{E}\right)\,\mathbf{I}-\mathbf{E} \left(\nabla^2 \mathbf{E}\right)\\
     &-\left(\nabla^2 \mathbf{E}\right)\mathbf{E}+\frac{1}{2}\left(\nabla\cdot \mathbf{E}\right)^2\, \mathbf{I} \Big],
    \end{split}
\end{equation}
as derived in the Supporing Information. While the above equation was derived for constant $\epsilon$ and $\ell_c$, the expression is identical if these parameters vary. For varying $\epsilon$ or $\ell_c$, the Korteweg-Helmholtz force density must be included in the electrostatic stress \cite{woodson1968electromechanical},
\begin{equation}
    \nabla \cdot \mathbf{\uptau}_e = \rho\mathbf{E}-\frac{1}{2}\mathbf{E}^2\nabla\epsilon+\frac{1}{2}(\nabla\cdot\mathbf{E})^2\nabla(\epsilon {\ell_c}^2).
\end{equation}

Within the distance of closest approach of the ions to the surface, correlations cannot affect the value of the Maxwell stress at the surface, $\mathbf{\uptau}_{e, \mathrm{surf}}$, generated by the surface charge density. The mechanical equilibrium problem therefore requires continuity in the electrostatic stress tensor evaluated in the solution and at the surface,
\begin{equation}\label{eq:eqBC}
    \mathbf{\uptau}_e-\mathbf{\uptau}_{e,\mathrm{surf}}=0.
\end{equation}
At a uniformly-charged, flat surface without a dielectric jump, the Maxwell stress tensor is simply $\mathbf{\uptau}_{e, \mathrm{surf}}={q_s}^2/(2\epsilon)\hat{n}\hat{n}$, while the Maxwell stress tensor in the electrolyte is given by Eq. \ref{eq:MaxStress}. Equating these two expressions, and substituting in $\hat{n}\cdot \mathbf{D}=q_s$ we arrive at a final boundary for a potential varying in one coordinate direction:
\begin{equation}
    \hat{n} \cdot \ell_c\nabla^3\phi=\nabla^2 \phi\Big\rvert_{S}
\end{equation}
applied at the distance of closest approach of the ion with the wall.

The method amounts to applying the Contact Theorem to the correlated electrolyte in the absence of correlations, shown below for $\mu_i^{\mathrm{ex}}=0$ at a flat electrode with constant charge density without a dielectric jump \cite{alawneh2007monte, henderson1979exact}:
\begin{equation} \label{eq:eqConstraint}
    P=-\frac{q_s^2}{2\epsilon}+kT\sum_i c_i\Big\rvert_{S}=-\hat{n}\cdot\mathbf{\uptau}_e\cdot\hat{n}+kT\sum_i c_i\Big\rvert_{S}.
\end{equation}
The Contact Theorem is a statement of mechanical equilibrium, where the repulsive osmotic pressure contribution is balanced by the electrostatic attraction from the Maxwell stress. Without the constraint from Eq. \ref{eq:eqBC}, the BSK theory does not obey this simple relationship which should be valid even for dilute electrolytes in the primitive model \cite{moreira2002simulations, burak2004test, carnie1981ionic}.
The procedure of ensuring continuity in the Maxwell stress can be repeated for any higher order $\hat\epsilon$ by equating the $\mathbf{\uptau}_e$ at successive orders of derivatives. The condition in Eq. \ref{eq:eqBC} is also applicable to any extended electrolyte mean-field theory with arbitrary models of concentrated solution activity and solvent polarizability, including interactions with a soft wall. The approach may even be extended to media with non-local dielectric constant $\hat{\epsilon}$ driven by solvent polarization \cite{kornyshev1978model, kornyshev1981nonlocal}.

\subsection{Correlation length scaling}
At a highly charged surfaces, the charge-charge correlations are dominated by the mutual repulsion of counterions at the interface, as demonstrated in the schematic in Fig. \ref{fig:fig1}(a). The size of a correlation hole of counterions forming a Wigner crystal is characterized by a length scale $R_\mathrm{hole}$:
\begin{equation}
    R_\mathrm{hole}=\left(\frac{1}{6}\right)^{1/4}\left(\frac{ze}{q_s}\right)^{1/2}
\end{equation}
For this work, we assume that the correlation length scales as the size of a correlation hole, determined by the surface charge denisty:
\begin{equation}
    \ell_c=\alpha R_\mathrm{hole}
\end{equation}
\begin{figure}
\centering
\includegraphics[width=0.4\linewidth]{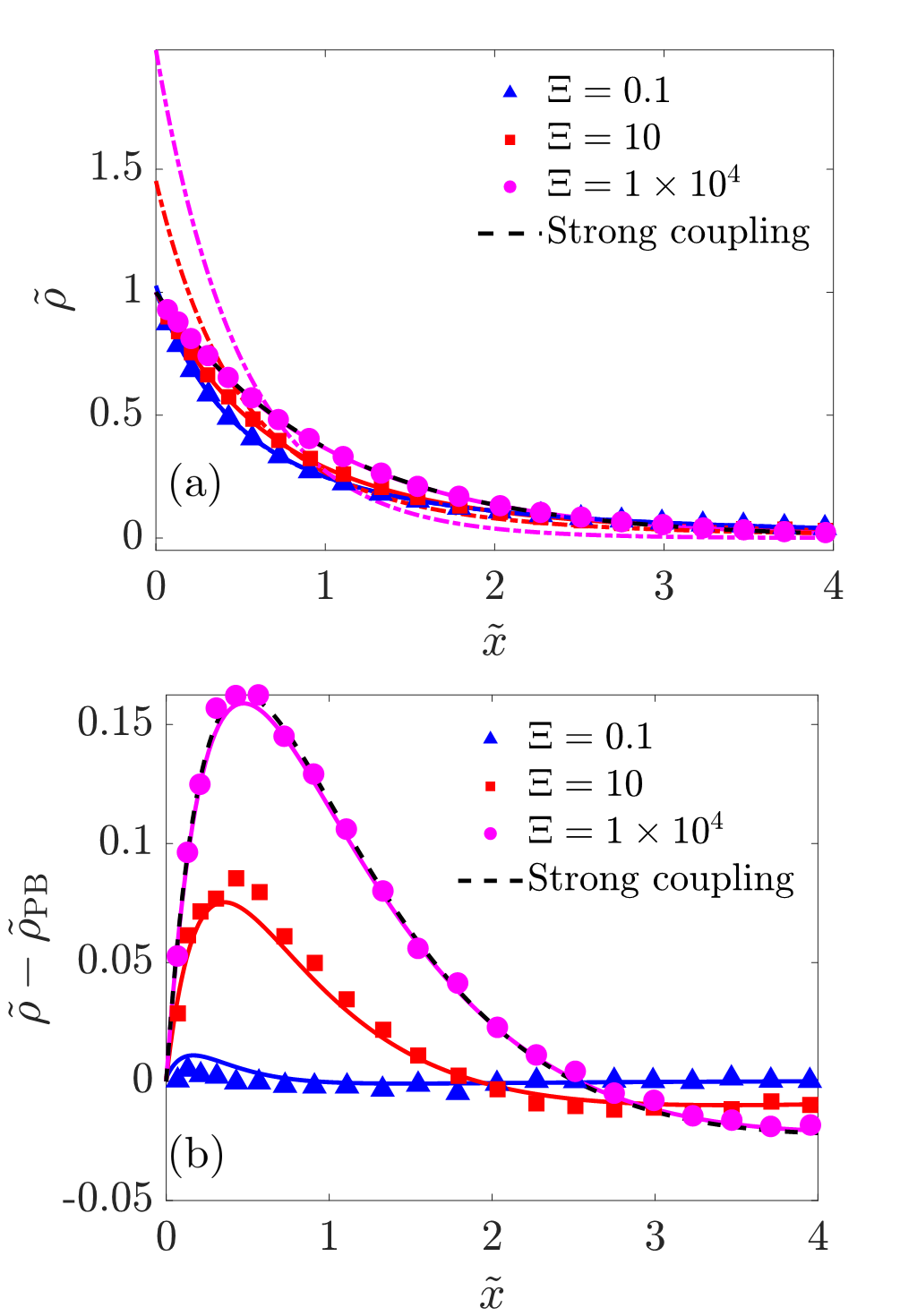}
\caption{BSK theory compared to MC simulations from \cite{moreira2002simulations} with $\alpha=0.50$ for counterions screening a charged isolated surface. The solid lines are the predictions of the BSK theory with the boundary condition of $\hat{n} \cdot \ell_c\nabla^3\phi=\nabla^2 \phi\Big\rvert_{S}$, the dashed-dotted lines are the predictions of the BSK theory with the boundary condition of $\hat{n} \cdot \ell_c\nabla^3\phi=0$ and the markers are from the MC simulations. Strong coupling limits are plotted as black dashed lines. (a) The charge density is plotted as a function of distance from an isolated surface. (b) The charge density difference relative to the solution to the PB theory as a function of distance from an isolated surface. }
\label{fig:fig2} 
\end{figure}
with one parameter $\alpha$, which is considered to be a constant. We will demonstrate that the fitted scaling with $\alpha=0.50$  works well from the limit of zero reservoir concentration (one-component plasma) to more concentrated electrolytes. 
At high concentration or at low surface charge densities, this scaling argument breaks down, and the other length scales might dominate in the correlation length. For example, if the surface charge tends to zero, then the charge-charge correlations will be dominated by the Bjerrum length and characteristic mean spacing between ions given by the bulk concentration. At very large charge densities and for large ion sizes, where $R_\mathrm{hole}$ becomes comparable to $a$, the ion diameter, the ion size can dominate in determining the charge-charge correlations due to over-crowding effects. Also, if other length scales are introduced, such as surface curvature, the correlation length between ions can also be affected, as demonstrated with the one-component plasma around a thin charged cylinder.

In the Supporting Information, the non-arbitrary scaling of the correlation length is investigated by comparison to Grand Canonical Monte Carlo simulations from ref. \cite{valisko2018systematic} for a $z:1$ electrolyte. Using the Buckingham-$\Pi$ theorem, we know that the dimensionless correlation length is related to a power law relationship, at least for small or large values of the dimensionless variables. Here we choose four length scales from which we can construct three dimensionless groups: $\ell_c$, $\ell_\mathrm{GC}$, $z^2\ell_B$, and $\lambda_D$, such that the powerlaw relationship can be expressed as:
 \begin{equation} \label{eq:eqBuckPi}
    \delta_c=\alpha_2\left(\frac{z^2\ell_B}{\ell_\mathrm{GC}}\right)^{\alpha_3}\left(\frac{z^2\ell_B}{\lambda_D}\right)^{\alpha_4},
\end{equation}

The correlation length scaling that arises from the fitting procedure to the MC data is given by:
\begin{equation}
    \ell_c\thicksim {\ell_B}^{1/4}(q_s/e)^{-1/8}{C_\mathrm{ref}}^{-1/6},
\end{equation}
with a fitted scale of:
\begin{equation} \label{eq:eqCor}
    \delta_c=0.35\left(\frac{z^2\ell_B}{\ell_\mathrm{GC}}\right)^{-1/8}\left(\frac{z^2\ell_B}{\lambda_D}\right)^{2/3}.
\end{equation}
Note that the fitted exponents are expressed in terms of fractions to emphasize their relationship to the intrinsic length scales in the system.

A detailed investigation of how correlations are affected as a function of surface charge density, ion valency, concentration, ion size, and surface curvature could motivate a more nuanced scaling of the correlation length, based on direct analysis of the charge-charge correlation function, including variations in the correlation length as a function of the distance from a charged surface. In this work, we isolate the electrostatic correlation effects for a dilute electrolyte at highly charged surfaces. For the purposes of the analysis in the main text, we will assume the correlation length to be $\ell_c=0.50 R_\mathrm{hole}$ for all the scenarios investigated in Fig. \ref{fig:fig1}. For the results with the fitted correlation length scaling in \ref{eq:eqCor}, one can refer to the Supporting Information.

\subsection{One-component plasma} 
 Considering a system of point-like counterions neutralizing a uniformly charged surface, the importance of correlations is governed by a coupling constant,
 \begin{equation}
     \Xi=2 \pi z^3 {\ell_B}^2 q_s/e, 
 \end{equation}
 which is a measure of the correlation hole size, $R_\mathrm{hole}$, compared to the characteristic ion distance from the surface, the Gouy-Chapman length,
 \begin{equation}
     \ell_\mathrm{GC}=e(2\pi z \ell_B q_s)^{-1}
 \end{equation}
such that $\Xi\sim {R_\mathrm{hole}}^2/{\ell_\mathrm{GC}}^2$. Here,  $z$ is the ion valency, $\ell_B$ is the Bjerrum length,  and $e$ is an elementary charge. In the weak coupling limit ($\Xi<<1$), PB theory is valid. In the strong coupling limit ($\Xi>>1$), counterions interact with the electric potential created by the surface since ion-surface interactions dominate \cite{moreira2002simulations, grosberg2002colloquium, burak2004test}.  

Now we consider applying the mechanical constraint, starting with the one-component plasma of infinitesimally small size with $\mu^{\mathrm{ex}}=0$. The one-component plasma consists of a single mobile ionic species which neutralizes the charge of a smeared out surface charge density. We can non-dimensionalize lengths with the Gouy-Chapman length, the potential by the thermal voltage for the counterion, $\phi_0=\frac{kT}{ze}$, and the charge density by $\rho_0=2\pi\ell_B q_s^2e^{-1}$. Here, we assume that the correlation length scales with the size of a correlation hole at the surface, $\delta_c=\ell_c/\ell_\mathrm{GC}=\alpha_0\sqrt{\Xi}$  Using $\thicksim$ to denote non-dimensionalized variables:
\begin{equation}\label{eq:eqOCP}
    {\alpha_0}^2 \Xi \tilde{\nabla}^4\tilde{\phi}-\tilde{\nabla}^2\tilde{\phi}=2\tilde\rho=2 e^{-\tilde{\phi}}
\end{equation}
with boundary conditions of
\begin{equation}
    \begin{split}
        &\hat n \cdot ({\alpha_0}^2\Xi\tilde{\nabla}^3\tilde\phi-\tilde\nabla\tilde\phi)=-2\\
        & \hat n \cdot \alpha_0\sqrt{\Xi}\tilde{\nabla}^3\tilde\phi=\tilde\nabla^2\tilde\phi
    \end{split}
\end{equation}
 at $\tilde x =0$, where $\alpha_0=1.36\alpha$ is an order one constant proportional to $\alpha$. Therefore, the importance of the higher order derivative is governed by the coupling parameter, $\Xi$.

The solution to these equations is compared to the results of MC simulations in Fig. \ref{fig:fig2} for a one-component plasma screening a plane of charge. The BSK theory reproduces the behavior of the one-component plasma from weak coupling, in intermediate coupling, and matches the strong coupling limit with $\alpha=0.50$. Furthermore, the BSK theory with the boundary condition of $\hat{n}\cdot\nabla^3\phi=0$,  represented by the dashed-dotted lines in Fig. \ref{fig:fig2}a, does not accurately represent the data at intermediate or strong coupling.

\begin{figure}
\centering
\includegraphics[width=0.4\linewidth]{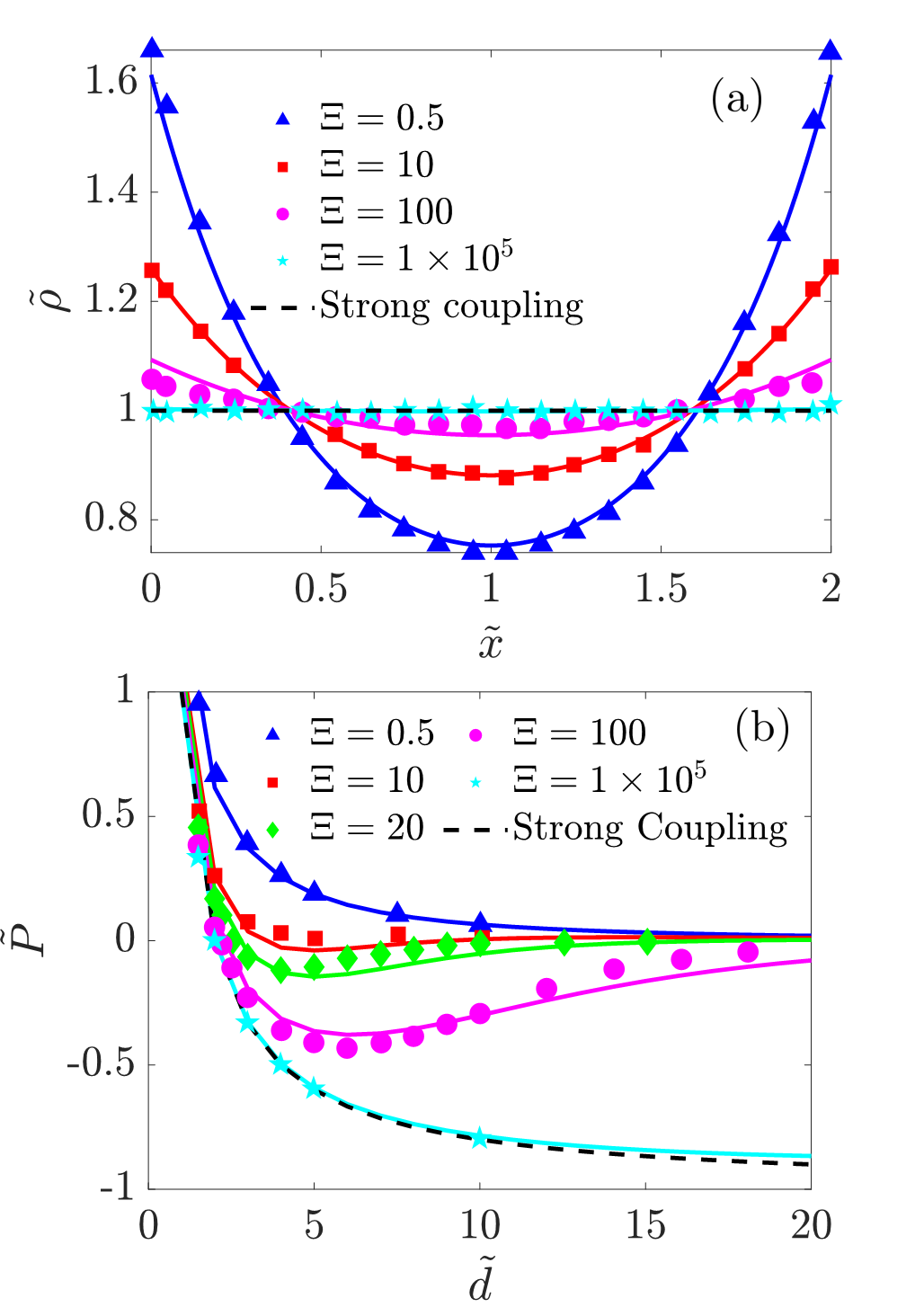}
\caption{BSK theory compared to MC simulations from \cite{moreira2002simulations} with $\alpha=0.50$ between two like-charged surfaces. The solid lines are the predictions of the BSK theory, and the markers are from the MC simulations. Strong coupling limits are plotted as dashed lines. (a) The charge density is plotted between two surfaces with separation $\tilde{d}=2$. (b) The pressure is calculated as a function of separation distance between the two plates. As the coupling increases, the pressures between the like-charged surfaces become attractive (negative) rather than repulsive (positive). The dimensionless pressure is $\tilde P= Pe^2/(2\pi\ell_B q_s^2)$. }
\label{fig:fig3} 
\end{figure}

We can also consider the one-component plasma between two charged surfaces of equal charge density with the same sign, confining the counterions in a gap of dimensionless distance $\tilde d$, as shown in Fig. \ref{fig:fig3}a. In Fig. \ref{fig:fig3}b, the pressure is plotted as a function of separation distances between two charged surfaces with different coupling parameters, using Eq. \ref{eq:eqBC} and using the same value for $\alpha$. The BSK theory again provides good agreement with the results of the MC simulations at all the coupling parameters. 

\begin{figure}
\centering
\includegraphics[width=0.4\linewidth]{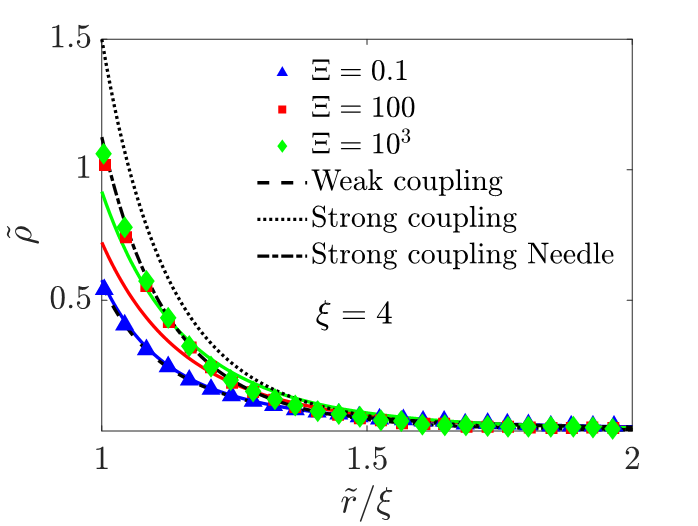}
\caption{BSK theory compared to MC simulations from \cite{mallarino2013counterion} using $\alpha=0.50$ for the counterion density around a charged cylinder for $\xi=4$. The solid lines are the results of applying Eq. \ref{eq:eqOCP} and the markers are the MC simulation results from \cite{mallarino2013counterion}.  The weak coupling, strong coupling, and re-normalized strong coupling needle limits are plotted \cite{mallarino2013counterion}. }
\label{fig:fig4} 
\end{figure}

Another critical question is the validity of Eq. \ref{eq:eqBC} at a curved interface. The simplest model system to test the hypothesis is the one-component plasma surrounding a charged cylinder of radius $\xi=R_{\mathrm{cyl}}/\ell_\mathrm{GC}$ at infinite dilution, corresponding to a cylindrical cell with outer radius $R_{\mathrm{out}}\rightarrow\infty$.  As shown in Fig. \ref{fig:fig4}, the BSK equation reproduces the results of the weak and strong coupling limits correctly. However, similar to the strong coupling expansion of \cite{naji2005counterions}, the theory does not correctly describe the renormalization of charge arising from Manning condensation in the needle limit, where a fraction $f=1-1/\xi$ of the charge is ``condensed" onto cylinder \cite{Manning1969}. The charge density must be multiplied by this fraction, $f$, in order to match the strong coupling expansion taking into account the charge renormalization in the needle limit \cite{mallarino2013counterion, cha2017hidden}. The smaller the radius of curvature, the more likely that the configuration of correlated ions is influenced by curvature. In the ``needle limit," where $\sqrt{\Xi}/\xi>>1$, ions are distributed in a nearly linear fashion along the cylindrical backbone with spacing scaling as $\tilde a\thicksim \Xi/\xi$, which may also be the relevant scaling for the correlation length in this regime rather than $\sqrt{\Xi}$. Supporting Figs. S1 and S2 show the results choosing $\delta_c=\Xi/\xi$, but the counterion condensation transition is still not captured.
\begin{figure} \label{fig:fig5}
\includegraphics[width=0.8\linewidth]{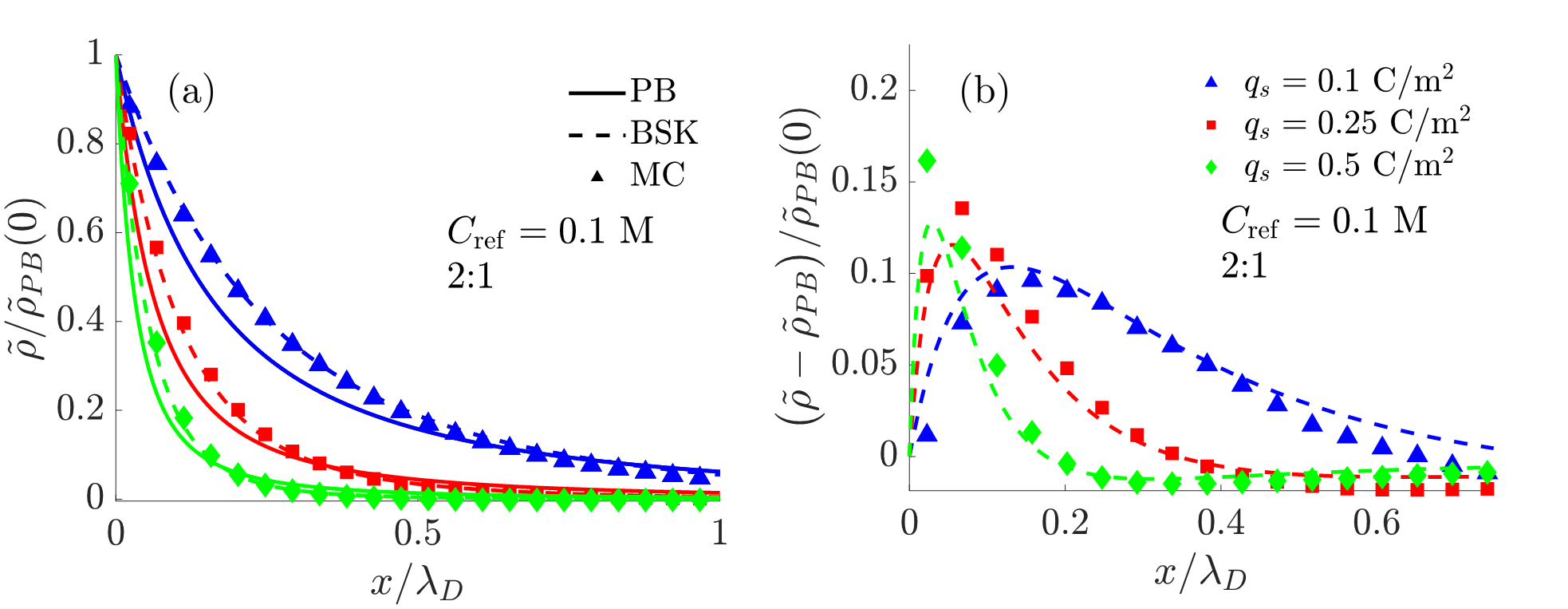}
\caption{BSK theory compared to MC simulations \cite{valisko2018systematic} of multivalent electrolytes with $\alpha=0.50$. (a) An example of the charge density profile for a 2:1 electrolyte at 0.1 M concentration compared to the GCMC simulations and PB theory. (b) The difference between the predictions of the BSK theory and the simulations from PB theory predictions. } 
\end{figure}
\section{Electrolytes}
A more useful and relevant application of the BSK theory is to describe the distribution of charges in electrolytes and ionic liquids, as was originally proposed. Here, we focus on the dilute electrolyte limit, to isolate electrostatic correlations directly, without complications from overcrowding.

If the BSK equation for a z:1 electrolyte with salt concentration ${C_\mathrm{ref}}$ is non-dimensionalized with the thermal voltage $\tilde{\phi}=(e\phi)/(kT)$ and the Debye length
\begin{equation}
    \lambda_D=\sqrt{\frac{\epsilon k T}{(z^2+z)e^2 {C_\mathrm{ref}}}},
\end{equation} 
$\tilde{\nabla}=\lambda_D\nabla$ and $\delta_c=\ell_c/\lambda_D$ the BSK equation becomes:
\begin{equation} \label{eq:eqElec}
    \delta_c^2 \tilde{\nabla}^4\tilde{\phi}-\tilde{\nabla}^2\tilde{\phi}=\tilde\rho=\frac{z e^{-z\tilde \phi}-z e^{\tilde \phi}}{z^2+z}.
\end{equation}
The boundary conditions are similarly modified to:
\begin{equation}
\begin{split}
    &\hat{n}\cdot(\delta_c^2\tilde{\nabla}^3\tilde{\phi}-\tilde{ \nabla}\tilde{\phi})=\tilde{q}_s \\
    &\hat{n}\cdot\delta_c\tilde{\nabla^3\phi}=\tilde\nabla^2\tilde\phi.
\end{split}
\end{equation}
The agreement of the predicted charge density profiles from Eq. \ref{eq:eqElec} with the GCMC data is very good, as exhibited in Fig. 4 for a 0.1 M 2:1 electrolyte. In the Supporting Information, the results are expanded to a more complete set of comparisons with simulations. It is seen with $\ell_c=0.50 R_\mathrm{hole}$ or with $\ell_c$ determined by Eq. \ref{eq:eqCor}, that the BSK theory can correct the PB charge density profiles, including an overscreening transition. Larger errors from the BSK theory are incurred at large concentration, where the current assumption of $\mu^{\mathrm{ex}}_i=0$ breaks down.

One implication of the boundary condition is that the differential capacitance for $\ell_c=0$ is equivalent to the case of $\ell_c\neq 0$ if $\mu_i^{\mathrm{ex}}=0$. Therefore, the differential capacitance for the correlated, dilute electrolyte is given by the traditional Gouy-Chapman equation:
\begin{equation}
    C_D=\frac{\epsilon}{\lambda_D}\cosh\left(\frac{\tilde{\phi}_D}{2}\right),
\end{equation}
in stark contrast to the original work in the limit of $\mu_i^{ex}=0$ \cite{Bazant2011, storey2012effects}. It would be interesting to explore the implications of the boundary condition on electrokinetic reversals, electrochemical interfaces, biological channels, or colloidal phenomena \cite{storey2012effects, kondrat2015dynamics,jiang2014dynamics, jiang2016current, huang2018confinement, Santos2009, stout2014continuum, moon2015osmotic, liu2013numerical, alijo2015effects, lee2013electric, alidoosti2018impact, mceldrew2018theory}. For example, the DLVO theory of colloidal interactions can be modified to include attractive correlation effects \cite{misra2019theory}.
\section{Further extensions of the theory}
\subsection{Charged Dielectric Interfaces:}
Note that the Maxwell stress condition (Eq. \ref{eq:MaxStress}) has only been stated without a dielectric jump. The stress condition may need further validation at a dielectric interface. A more general statement of matching the Maxwell stress with and without correlations might be given by a jump condition between the two media:
\begin{equation}
    \hat{n}\cdot[\mathbf{\uptau}_{e,1}-\mathbf{\uptau}_{e,2}]=\hat{n}\cdot[\mathbf{\uptau}_{e,1}-\mathbf{\uptau}_{e,2}]_{\ell_c=0}.
\end{equation}
 For a uniformly-charged, flat interface without a dielectric jump, $\tau_{e,2}=0$, so the RHS of the above equation reduces to $\hat{n}\cdot[\mathbf{\uptau}_{e,1}-\mathbf{\uptau}_{e,2}]_{\ell_c=0}=\frac{{q_s}^2}{2\epsilon_1}\hat{n}$ .

\subsection{Concentrated Electrolytes and Ionic Liquids:}
The present analysis attempts to isolate the effect of ion correlations in a dilute electrolyte.  Ion size effects, particularly for $a/\lambda_D>1$ will require further validation to properly account for correlations guided by ion size combined with electrostatics. A non-local free energy functional might be necessary to capture the size correlations in concentrated solutions \cite{roth2010fundamental,gillespie2015review}, in conjunction with electrostatic correlations. Short-range bulk correlations \cite{goodwin2017mean} are not captured in this theory. Furthermore, if surface charges are discrete rather than smeared out, the contact condition may change \cite{moreira2002counterions2}. For an arbitrary mixture of ions with different valency, the effective correlation length will depend upon correlations between each pair of species, although the correlations at high surface charge density will still be dominated by the most highly charged counterion.

\section{Conclusions}
The phenomenological BSK theory describes non-local, discrete correlation effects with a higher-order, local, continuum description of the free energy quite well.  The remarkable agreement of the theory with the one-component plasma and primitive model electrolyte suggest that higher-order, continuum equations can properly account for correlation effects, as long as the appropriate constraints are imposed at boundaries.  The formalism used here could be extended to the ionic liquid limit, although ion pairing, short-range non-electrostatic correlations, and ``spin glass" ordering \cite{levy2018spin} might preclude a simple continuum description. Further modifications are needed to capture the long range screening exhibited in ionic liquids and concentrated electrolytes\cite{smith2016electrostatic}, as well as density oscillations expected in overcrowded systems \cite{roth2010fundamental}. Even so, the BSK theory captures important features of electrostatic correlations, including like-charge attraction and overscreening, driven by electrostatic interactions of spatially correlated counterions. Furthermore, unlike many previous approaches, all the electrostatic forces are contained self-consistently within the electrostatic potential, $\phi$. Detailed analysis of experimental data is needed to determine the competing effects of surface adsorption reactions modifying fixed surface charge \cite{mugele2015ion} or the overscreening/ like-charge attraction effects modeled by the BSK theory \cite{misra2019theory}.

\section{Acknowledgements}
This work  was supported by an Amar G. Bose Research Grant (electrolyte calculations) and the Center for Enhanced Nanofluidic Transport (one-component plasma calculations). JPD is also supported by the National Science Foundation Graduate Research Fellowship under Grant No. 1122374. JPD would like to acknowledge useful discussions with Michael McEldrew, Amir Levy, Tingtao Zhou, Dimitrios Fraggedakis, and Mohamed Mirzadeh.

\bibliography{library}

\end{document}


\maketitle

\section{Derivation of Maxwell Stress}
Starting from the equation
\begin{equation}
    \nabla \cdot \mathbf{\uptau}_e = \rho\mathbf{E}= \nabla \cdot(\hat\epsilon\mathbf{E})\mathbf{E},
\end{equation}
the derivation of the Maxwell stress makes use of two vector identities:
\begin{equation}
    \begin{split}
        &\nabla(\mathbf{A}\cdot\mathbf{B})=\mathbf{A}\cdot \nabla \mathbf{B}+\mathbf{B}\cdot \nabla \mathbf{A}\\
        &\nabla\cdot(\mathbf{A}\mathbf{B})=(\nabla \cdot \mathbf{A})\mathbf{B}+\mathbf{A}\cdot\nabla\mathbf{B},
    \end{split}
\end{equation}
for two vectors $\mathbf{A}$ and $\mathbf{B}$, valid when $\nabla\times\mathbf{A}=0$ and $\nabla\times\mathbf{B}=0$.
Using the permittivity operator $\hat\epsilon=\epsilon(1-{\ell_c}^2\nabla^2)$, the expression can be split into two terms that can be analyzed separately, 
\begin{equation}
    \nabla \cdot \mathbf{\uptau}_e = \nabla\cdot(\epsilon \mathbf{E})\mathbf{E}-\nabla\cdot(\epsilon{\ell_c}^2\nabla^2\mathbf{E})\mathbf{E}.
\end{equation}
With the requirement of $\nabla\times\mathbf{E}=0$, the first term can be written as 
\begin{equation}
    \begin{split}
    \nabla\cdot(\epsilon \mathbf{E})\mathbf{E}=&\nabla\cdot(\epsilon \mathbf{E})\mathbf{E}+\frac{1}{2}\epsilon\mathbf{E}\cdot\nabla\mathbf{E}+\frac{1}{2}\mathbf{E}\cdot\nabla(\epsilon\mathbf{E})-\frac{1}{2}(\epsilon\mathbf{E}\cdot\mathbf{E})\\
    =&\nabla\cdot(\epsilon \mathbf{E})\mathbf{E}+\epsilon\mathbf{E}\cdot\nabla\mathbf{E}+\frac{1}{2}\mathbf{E}\cdot\mathbf{E}\nabla\epsilon-\frac{1}{2}(\epsilon\mathbf{E}\cdot\mathbf{E})\\
    =&\nabla\cdot(\epsilon \mathbf{E E}-\frac{1}{2}\epsilon\mathbf{E}^2\,\mathbf{I})+\frac{1}{2}\mathbf{E}\cdot\mathbf{E}\nabla\epsilon\\
    =&\nabla\cdot(\epsilon \mathbf{E E}-\frac{1}{2}\epsilon\mathbf{E}^2\,\mathbf{I}),
    \end{split}
\end{equation}
with constant $\epsilon$, where $\mathbf{I}$ is the identity tensor. The second term can be similarly expanded:
\begin{equation}
    \begin{split}
        -\nabla\cdot(\epsilon{\ell_c}^2\nabla^2\mathbf{E})\mathbf{E}=&-\nabla\cdot(\epsilon{\ell_c}^2\nabla^2\mathbf{E})\mathbf{E}-\epsilon{\ell_c}^2\nabla^2\mathbf{E}\cdot\nabla\mathbf{E}\\
        &-\mathbf{E}\cdot \nabla(\epsilon{\ell_c}^2\nabla^2\mathbf{E})+\nabla(\mathbf{E}\cdot\epsilon{\ell_c}^2\nabla^2\mathbf{E})\\
        =-\nabla\cdot(\epsilon{\ell_c}^2\nabla^2\mathbf{E})\mathbf{E}&-\epsilon{\ell_c}^2\nabla^2\mathbf{E}\cdot\nabla\mathbf{E}
        -\mathbf{E}\cdot \nabla(\epsilon{\ell_c}^2\nabla^2\mathbf{E})\\+\nabla(\mathbf{E}\cdot\epsilon{\ell_c}^2\nabla^2\mathbf{E})& +\mathbf{E}\cdot\nabla(\epsilon{\ell_c}^2\nabla^2\mathbf{E})+(\nabla\cdot\mathbf{E})\epsilon{\ell_c}^2\nabla^2\mathbf{E}\\
        -\nabla\cdot(\mathbf{E}\epsilon{\ell_c}^2\nabla^2\mathbf{E})&\\
        =\nabla\cdot\big[-\epsilon{\ell_c}^2(\nabla^2\mathbf{E})&\mathbf{E}-\epsilon {\ell_c}^2\mathbf{E}(\nabla^2\mathbf{E}) 
        +\mathbf{E}\cdot(\epsilon{\ell_c}^2\nabla^2\mathbf{E})\mathbf{I}\big]\\+(\nabla\cdot\mathbf{E})\epsilon{\ell_c}^2\nabla^2\mathbf{E}&.
    \end{split}
\end{equation}
At this point, another product rule must be used:
\begin{equation}
    a\nabla a=\frac{1}{2}\nabla(a^2),
\end{equation}
where $a$ is a scalar. Applying the above identity gives:
\begin{equation}
    \begin{split}
        -\nabla\cdot(\epsilon{\ell_c}^2\nabla^2\mathbf{E})\mathbf{E}=&\\
        \nabla\cdot\big[-\epsilon{\ell_c}^2(\nabla^2\mathbf{E})&\mathbf{E}-\epsilon {\ell_c}^2\mathbf{E}(\nabla^2\mathbf{E}) 
        +\mathbf{E}\cdot(\epsilon{\ell_c}^2\nabla^2\mathbf{E})\mathbf{I}\big]\\+\frac{\epsilon{\ell_c}^2}{2}\nabla\left((\nabla\cdot\mathbf{E})^2\right)
        \\
        =\nabla\cdot\big[-\epsilon{\ell_c}^2(\nabla^2\mathbf{E})&\mathbf{E}-\epsilon {\ell_c}^2\mathbf{E}(\nabla^2\mathbf{E}) 
        +\mathbf{E}\cdot(\epsilon{\ell_c}^2\nabla^2\mathbf{E})\mathbf{I}\big]\\+\nabla\cdot\left[\frac{\epsilon{\ell_c}^2}{2}(\nabla\cdot\mathbf{E})^2\mathbf{I}\right]&-\frac{1}{2}(\nabla\cdot\mathbf{E})^2\nabla(\epsilon{\ell_c}^2)
        \\
        =\nabla\cdot\big[-\epsilon{\ell_c}^2(\nabla^2\mathbf{E})&\mathbf{E}-\epsilon {\ell_c}^2\mathbf{E}(\nabla^2\mathbf{E}) 
        +\mathbf{E}\cdot(\epsilon{\ell_c}^2\nabla^2\mathbf{E})\mathbf{I}\big]\\+\nabla\cdot\left[\frac{\epsilon{\ell_c}^2}{2}(\nabla\cdot\mathbf{E})^2\mathbf{I}\right]&,
    \end{split}
\end{equation}
with constant $\epsilon$ and $\ell_c$. The final expression for the Maxwell stress is therefore:
\begin{equation}
\begin{split}
    \mathbf{\uptau}_e=& \epsilon \mathbf{E E}-\frac{1}{2}\epsilon\mathbf{E}^2\,\mathbf{I} 
     +\epsilon {\ell_c}^2\Big[\left(\mathbf{E}\cdot \nabla^2 \mathbf{E}\right)\,\mathbf{I}-\mathbf{E} \left(\nabla^2 \mathbf{E}\right)\\
     &-\left(\nabla^2 \mathbf{E}\right)\mathbf{E}+\frac{1}{2}\left(\nabla\cdot \mathbf{E}\right)^2\, \mathbf{I} \Big].
    \end{split}
\end{equation}

 
\newpage
\section{One-component plasma around a cylinder}
\begin{figure}[htb]
\centering
\includegraphics[width=0.4\linewidth]{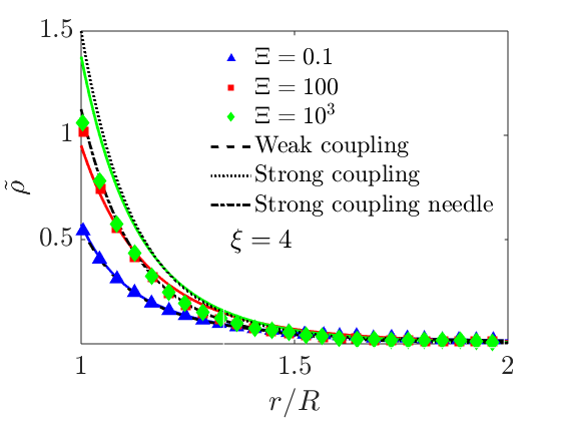}
\caption{Re-plotted version of Fig. 3 in the main text using a different correlation length scaling. BSK theory compared to additional MC simulations from \cite{mallarino2013counterion} using $\delta_c= \Xi/\xi$ for the counterion density around a charged cylinder for $\xi=4$. The labels are identical to Fig. 3 in the main text. }
\label{fig:figS1} 
\end{figure}

\begin{figure}[htb]
\centering
\includegraphics[width=0.4\linewidth]{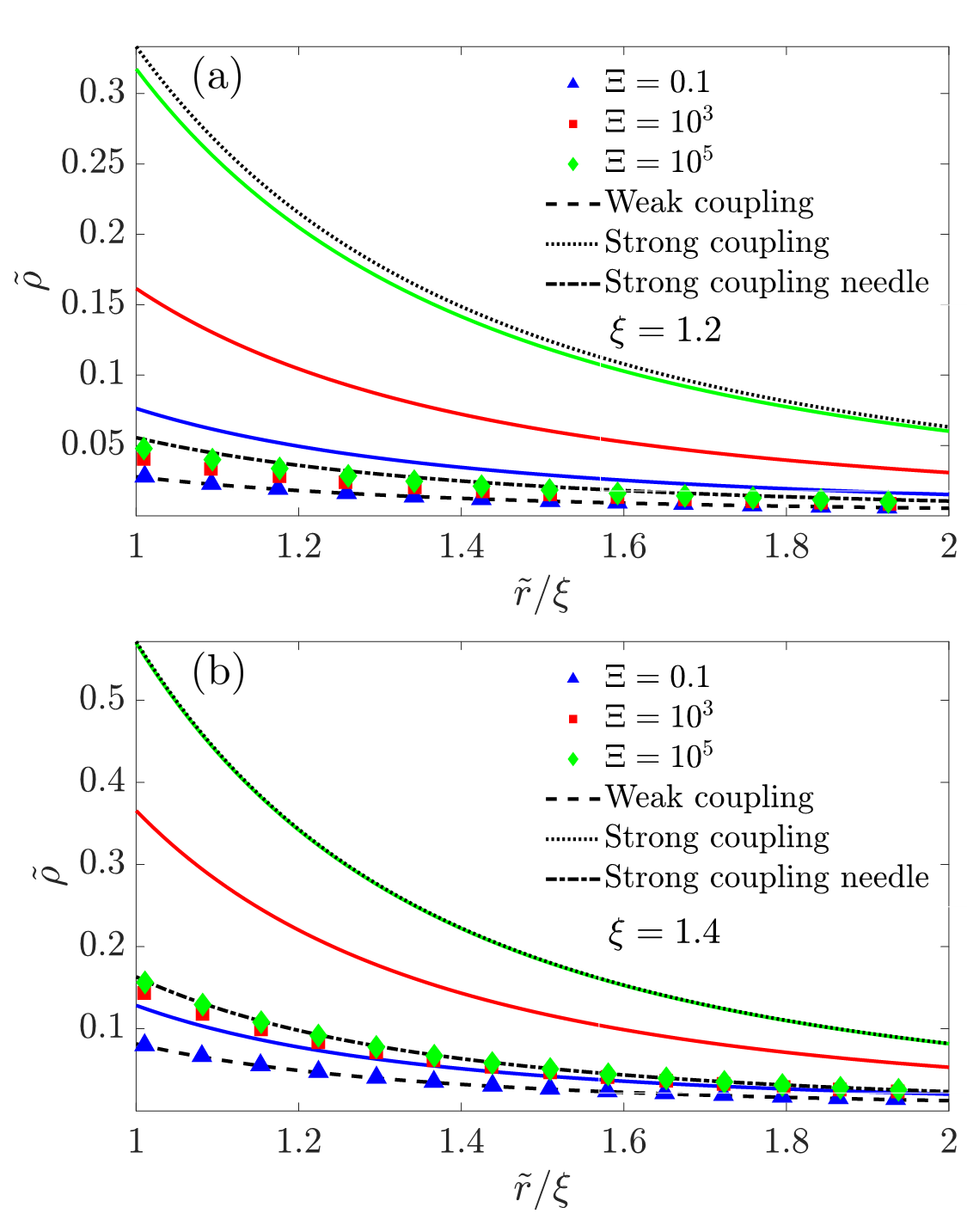}
\caption{BSK theory compared to additional MC simulations from \cite{mallarino2013counterion} using $\delta_c= \Xi/\xi$ for the counterion density around a charged cylinder for (a) $\xi=1.2$ and  (b) $\xi=1.4$. The labels are identical to Fig. 3 in the main text.}
\label{fig:figS2} 
\end{figure}

Additional results for the one-component plasma around a cylinder are exhibited in Figs. \ref{fig:figS1} and \ref{fig:figS2}, using the correlation length scaling $\delta_c=\Xi/\xi$ (the correlation hole size in the needle limit \cite{mallarino2013counterion}). The figures show the occurrence of Manning criticality in MC simulations, which cannot be reproduced by the BSK theory over the finite domain over which the equations are being solved. Even so, the BSK theory can reproduce the transition to the strong coupling limit. While the condensation phenomenon should only occur at infinite dilution, a finite system size must be chosen for the numerical solution of the BSK equation. The parameter $\Delta=\ln(R_\mathrm{out}/R_\mathrm{cyl})$ is varied between 6.2 and 13 to ensure numerical accuracy of the solutions over the large domain, depending on the value of $\delta_c$.
\newpage
\section{Comparison to full electrolyte data set}
\begin{figure*}[b] 
\includegraphics[width=0.8\linewidth]{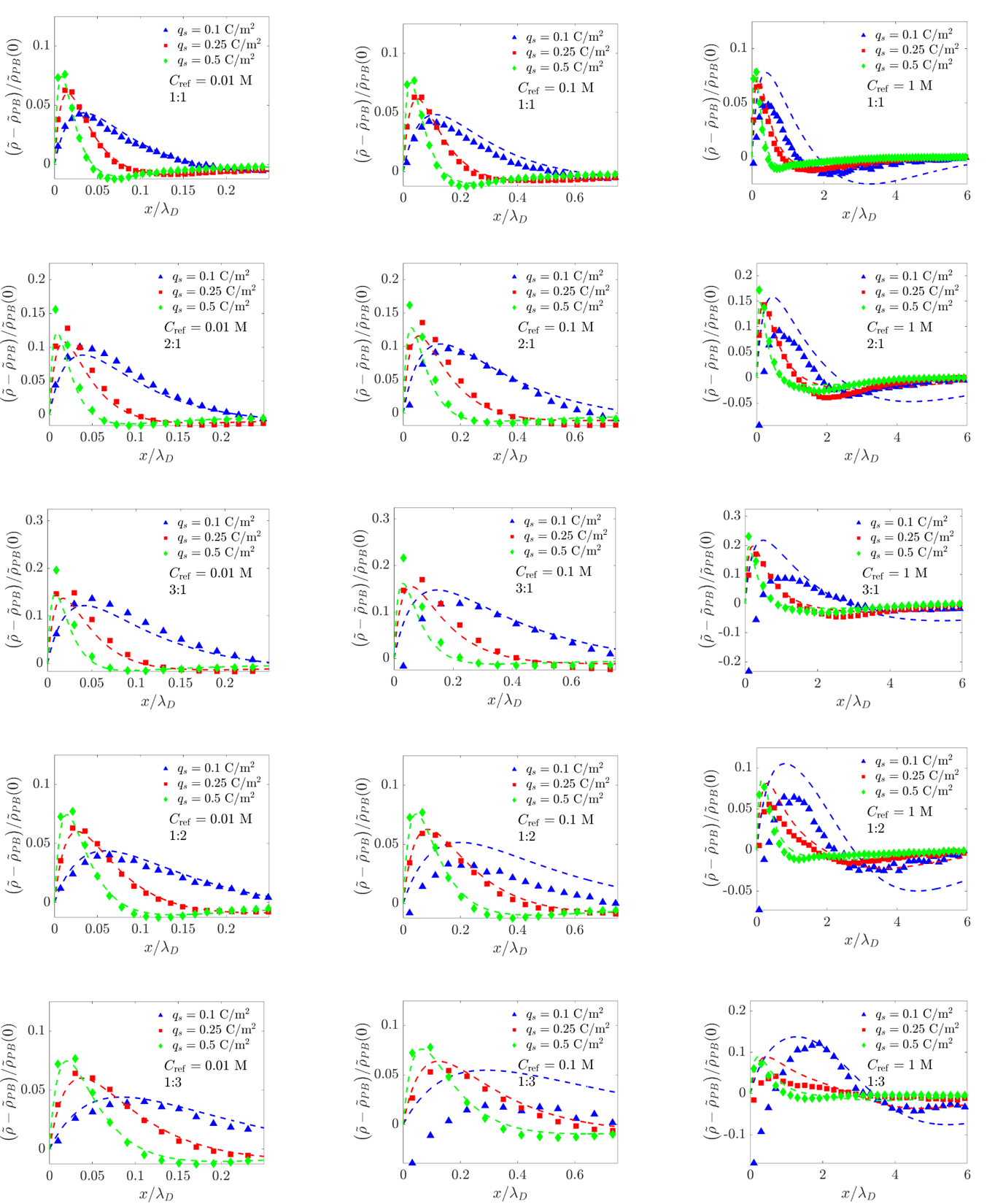}
\caption{Comprehensive comparison of BSK theory with MC simulations from \cite{valisko2018systematic} for the correlation length scaling $\ell_c=0.50 R_\mathrm{hole}$ plotted as a difference compared to PB theory. Note that the profiles are organized by $z_\mathrm{counterion}:z_\mathrm{coion}$ in each row and by $C_\mathrm{ref}$ in each column.}
\label{fig:figS3} 
\end{figure*}

\begin{figure*}[b] 
\includegraphics[width=0.8\linewidth]{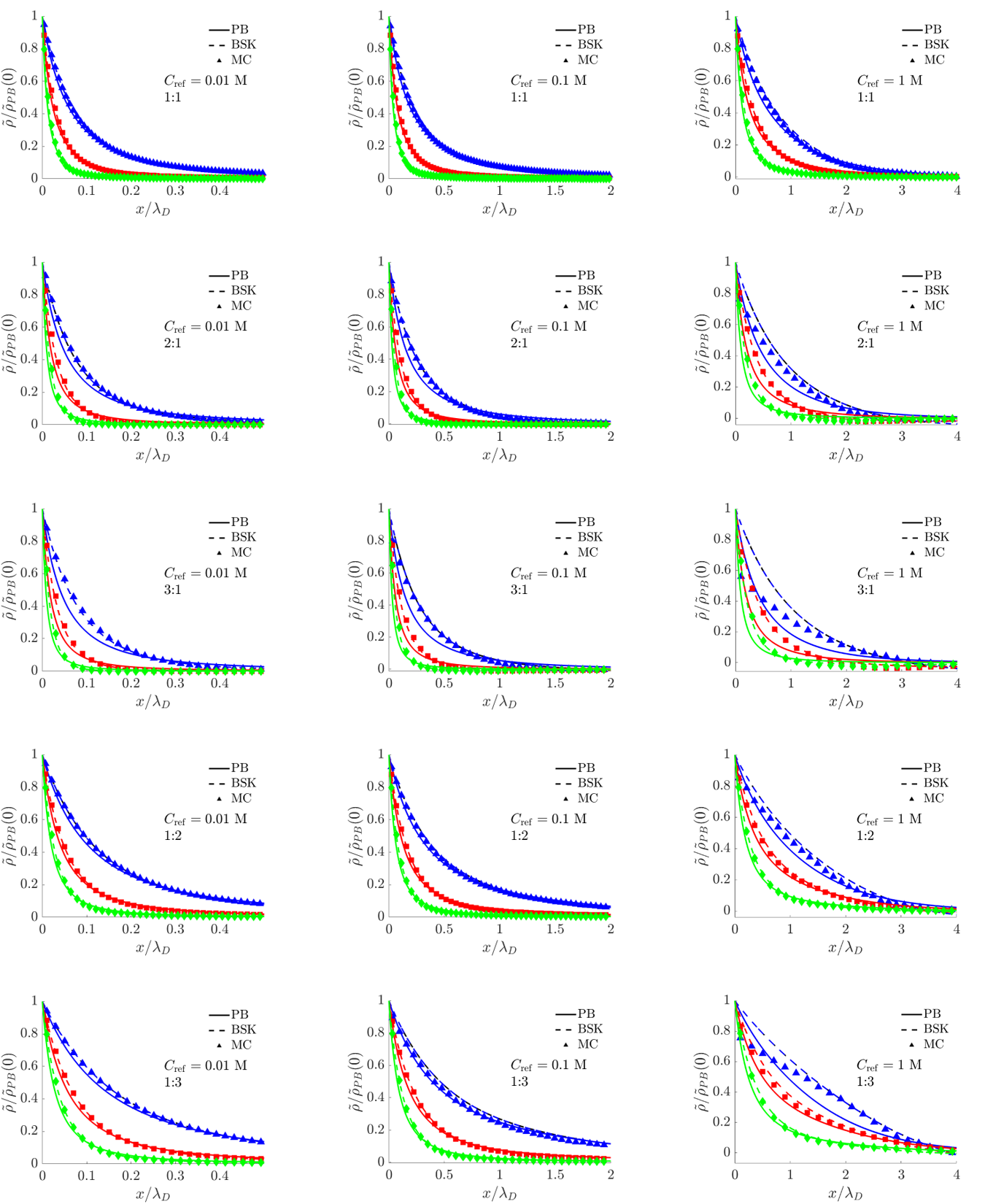}
\caption{Comprehensive comparison of BSK theory with MC simulations from \cite{valisko2018systematic} for the correlation length scaling $\ell_c=0.50 R_\mathrm{hole}$ plotted as the total charge density. Note that the profiles are organized by $z_\mathrm{counterion}:z_\mathrm{coion}$ in each row and by $C_\mathrm{ref}$ in each column.}
\label{fig:figS4} 

\end{figure*}

\begin{figure*}[b] 
\includegraphics[width=0.8\linewidth]{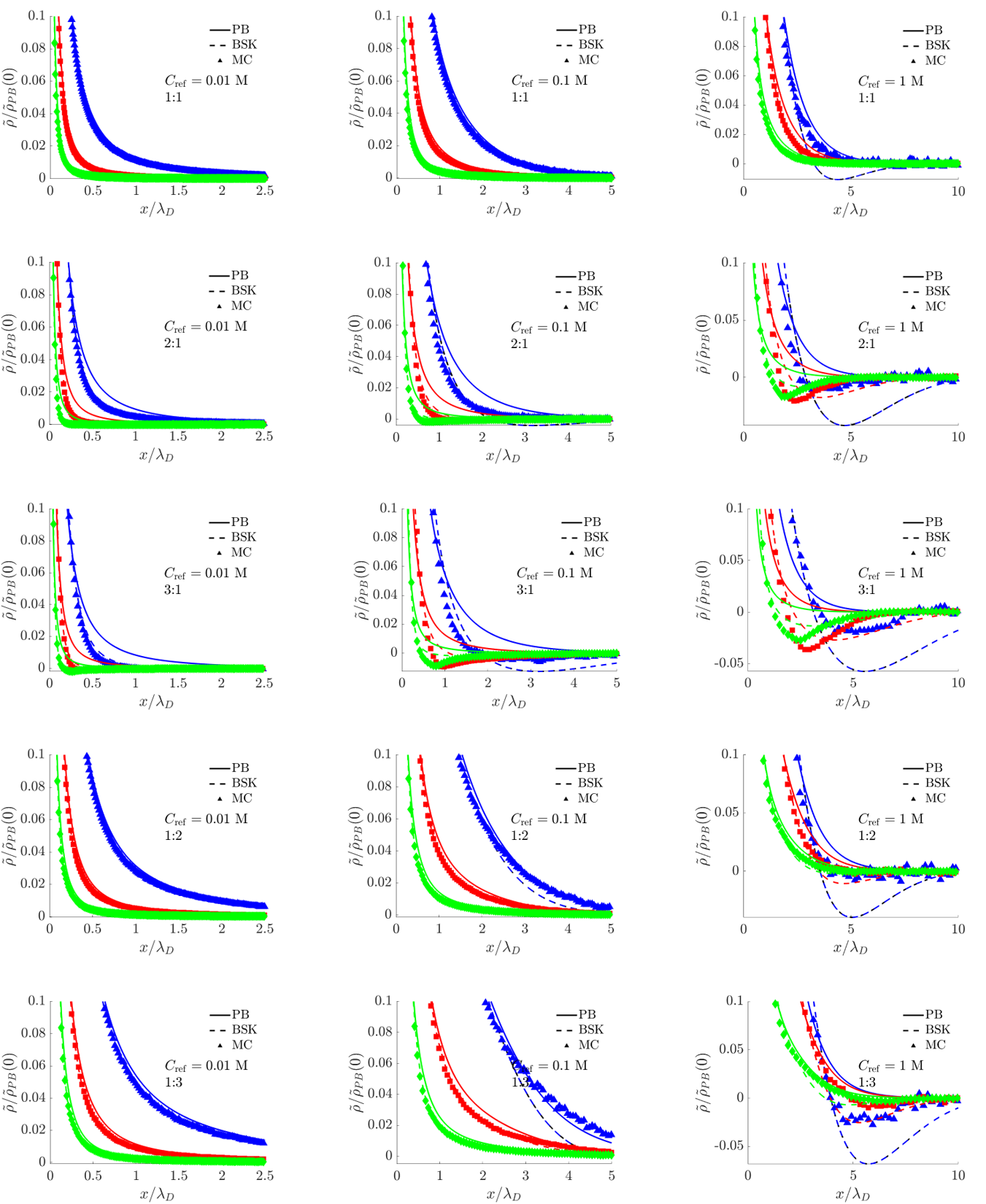}
\caption{Comprehensive comparison of BSK theory with MC simulations from \cite{valisko2018systematic}for the correlation length scaling $\ell_c=0.50 R_\mathrm{hole}$ plotted as the total charge density. The axis are zoomed in to isolate the extent of overscreening. Note that the profiles are organized by $z_\mathrm{counterion}:z_\mathrm{coion}$ in each row and by $C_\mathrm{ref}$ in each column.}
\label{fig:figS5} 
\end{figure*}

The comparisons the data from \cite{valisko2018systematic} is explored more extensively in Figs. \ref{fig:figS3}, \ref{fig:figS4}, and \ref{fig:figS5}  for $\ell_c=0.50 R_\mathrm{hole}$. Respectively, they show the difference of the BSK theory compared to PB theory, the full charge density profiles, and zoomed in profiles into the overscreening region. The BSK theory can closely match the structure of the charge density for many of the plots, including predicting the occurrence of overscreening, with the same correlation length scaling as the one-component plasma. However, at large concentrations, the excess electrochemical potential and ion size effects play a bigger role, leading to inaccuracies of the theory. Even so, the theory is quite adequate up to to provide the correct qualitative corrections up to 1M for the symmetric ions. Note that the MC profiles are shifted by one ion radius, so that the zero $x$ values of the theory and the simulations match. In other words, only the diffuse layer charge density is plotted.

\section{Empirical correlation length scaling in electrolytes}
 Given the large amount of simulation data on the restricted primitive model, it is possible to leverage the simulation data to fit the correlation length scaling. To define a correlation length that could be used uniformly across many different charge densities, concentrations, and valencies, we used the Grand Canonical MC data from \cite{valisko2018systematic}, and fit the correlation length by matching to the charge density profile from an isolated surface with $\epsilon=78.5\epsilon_0$ for a range of conditions: 1:1, 2:1, 3:1, 1:2, and 1:3 ($z_\mathrm{counterion}:z_\mathrm{coion}$) electrolytes, ${C_\mathrm{ref}}$ of 0.01 M, 0.1 M, and 1 M, and $q_s$ of 0.02, 0.04, 0.06, 0.08, 0.1, 0.175, 0.25, 0.375, and 0.5 C/m$^2$. The finite size of ions with symmetric 0.3 nm diameter is not taken into account in solving the BSK equation. Instead, only the diffuse part of the double layer is plotted, so that the $x$-axis is shifted by one ionic radius. In some cases, the contact densities for the simulations do not match PB, particularly at higher concentrations and lower charge densities, due to problems with the current assumption of $\mu_i^{\mathrm{ex}}=0$. In these cases, the profiles cannot be adequately fit by varying the correlation length, and this leads to large errors in the fits. We choose not to add more complexity to the model to capture these exceptional data points, and to rather focus on the electrostatic correlation component. Therefore, we distinguish which fitted values of the correlation length correspond to good fits and which correspond to bad fits. The sum of square error of the fit between the BSK solution and the MC data in units of the PB contact density is chosen as a metric to distinguish between good fits and bad fits. Good fits are defined as those having sum of square error $<0.005$.
 
\begin{figure}[htb]
\centering
\includegraphics[width=1\linewidth]{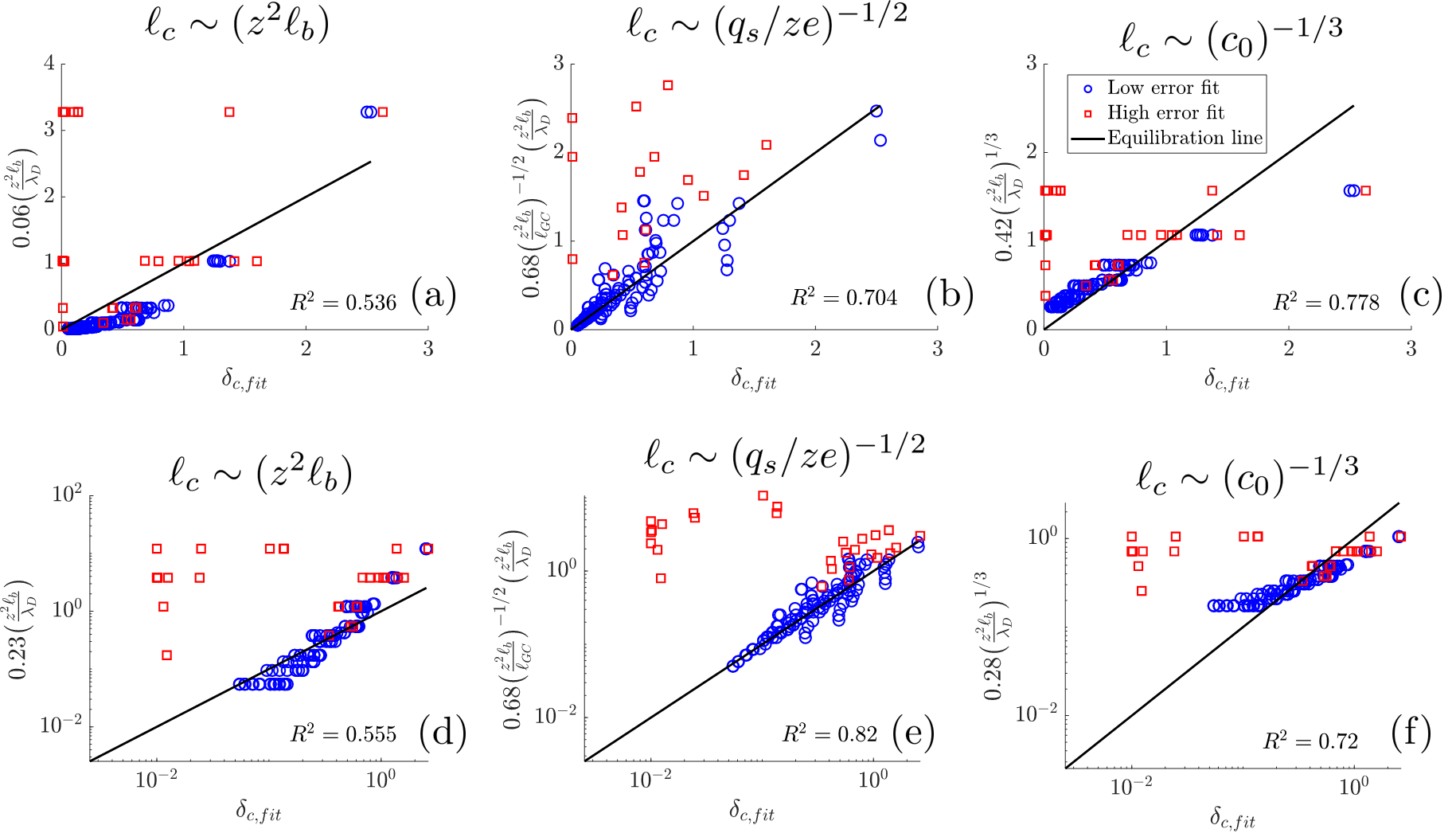}
\caption{Relating the correlation length to individual length scales in the system.  Fitted $\delta_c$ values for all simulations are plotted on the $x$-axis versus the $\delta_c$ given by $\delta_c=\ell_c/\lambda_D=\alpha_i \ell_i/\lambda_D$, where $\ell_i=z^2 \ell_b$ (a,d), $\ell_i=(q_s/(ze))^{-1/2}$ (b,e), and $\ell_i=c_0^{-1/3}$ (c,f). The top row is plotted on a linear scale, whereas the bottom row is plotted on a log scale. Note that the value of $\alpha_i$ is fit for a, c, d, and f, but is fixed for b and e based on the value determined for the one component plasma ($\ell_c=0.50 R_\mathrm{hole}$). }
\label{fig:figS6} 
\end{figure}

 We can then use the fitted correlation lengths with low error to relate the correlation length to the intrinsic length scales in the system. For example, we can fit the constant $\alpha_1$ if we assume that the correlation length scales with the  $\ell_c=\alpha_1 z^2 l_b$. We can choose $\alpha_1$ so as to minimize the error between the fitted correlation lengths with low error and the predicted one that is proportional to the Bjerrum length. However, if we fit the correlation length to be proportional to an individual length scale in the system, we see that the fitted correlation length has some dependence on other length scales in the system. For example, we can choose the correlation length scale determined from the one-component plasma, $\ell_c=0.50 R_\mathrm{hole}$, and compare it to the fitted correlation lengths for the electrolyte data set. We find that there is some concentration dependencies, as shown in Figs. \ref{fig:figS6}b and \ref{fig:figS6}e.  A more extensive comparison of fitting individual correlation lengths is shown in Fig. \ref{fig:figS6}. One can also observe that the definition of the correlation length based on the Bjerrum length $\ell_c\sim z^2 l_b$ is not the appropriate choice for the correlation length based on the poor agreement between the fitted values and the predicted ones.
 
\begin{figure}[htb]
\centering
\includegraphics[width=0.8\linewidth]{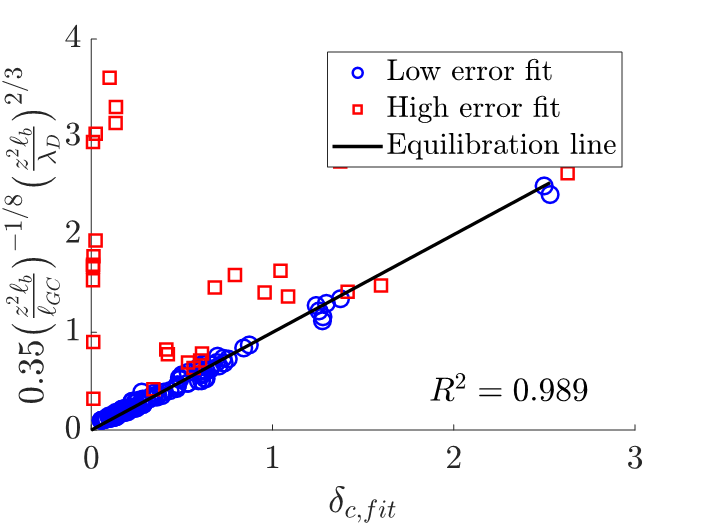}
\caption{The agreement of
the fitted correlation lengths with the scaling from Eq. \ref{eq:eqCor}. Fitted $\delta_c$ values for all simulations are plotted on the $x$-axis versus the $\delta_c$ given by Eq. \ref{eq:eqCor}. The profiles that can be fit with low error are used to determine the fitted scaling (marked in blue).}
\label{fig:figS7} 
\end{figure}

 Because of the dependence on all length scales, Using only profiles that can be fit with low error, we fit the fitted correlation length, $\delta_{c,\mathrm{fit}}$, to a power law relationship of dimensionless quantities. 
 \begin{equation} \label{eq:eqBuckPi}
    \delta_c=\alpha_2\left(\frac{z^2\ell_B}{\ell_\mathrm{GC}}\right)^{\alpha_3}\left(\frac{z^2\ell_B}{\lambda_D}\right)^{\alpha_4},
\end{equation}
 The result is given in Fig. \ref{fig:figS7} and below:
\begin{equation} \label{eq:eqCor}
    \delta_c=0.35\left(\frac{z^2\ell_B}{\ell_\mathrm{GC}}\right)^{-1/8}\left(\frac{z^2\ell_B}{\lambda_D}\right)^{2/3},
\end{equation}
where $z$ is the counterion valency. This relationship gives the scaling
$\ell_c\thicksim {\ell_B}^{1/4}(q_s/e)^{-1/8}{C_\mathrm{ref}}^{-1/6}$.
The correlation length scaling is thus a combination of the intrinsic lengths in the system, $\ell_B$, $(q_s/e)^{-1/2}$, and ${C_\mathrm{ref}}^{-1/3}$. Note that the fitted quantities are represented as fractions so as to emphasize their dependence on the fundamental length scales of the system.

\begin{figure*}[b] 
\includegraphics[width=0.8\linewidth]{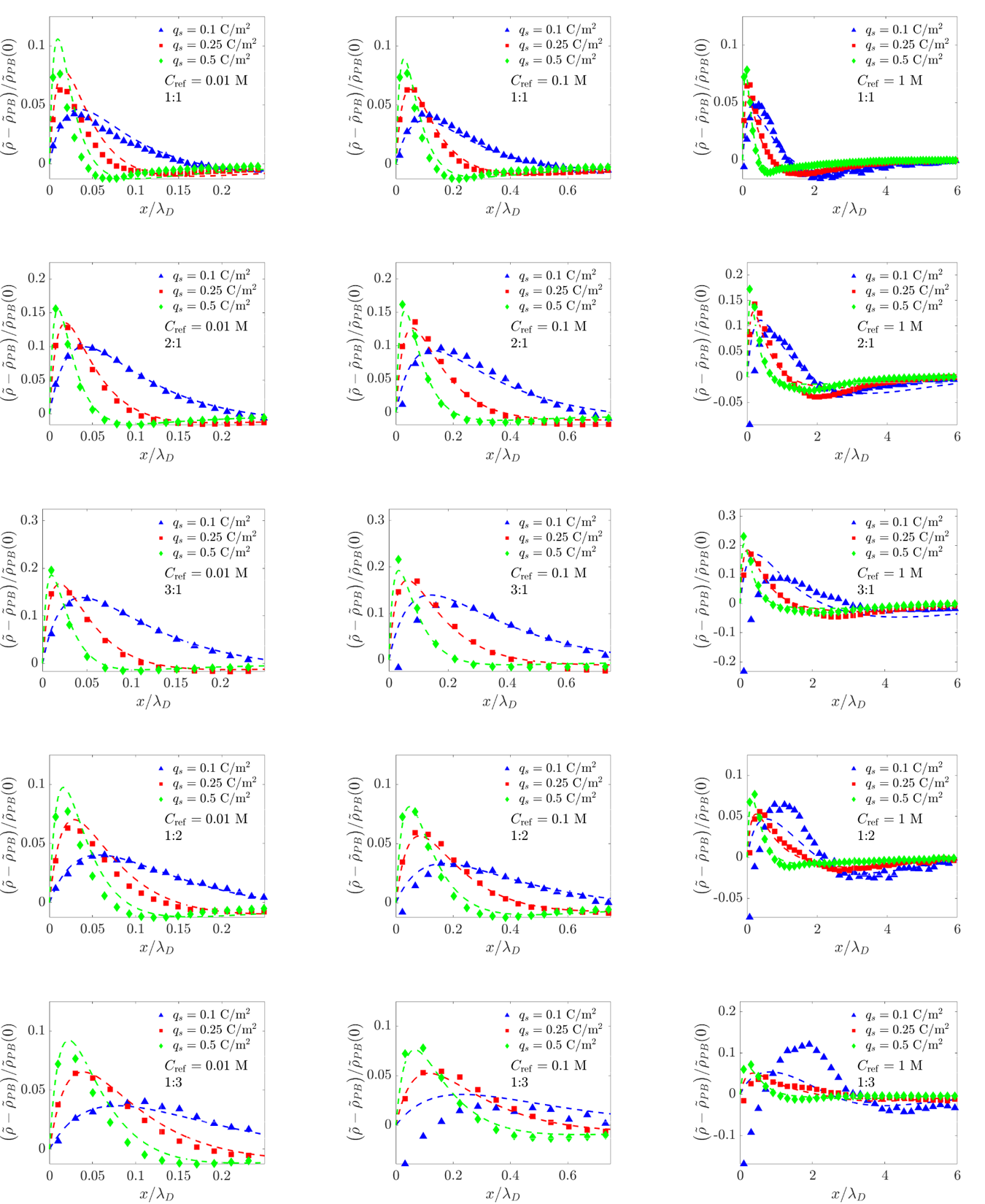}
\caption{Comprehensive comparison of BSK theory with MC simulations from \cite{valisko2018systematic} for the correlation length scaling from Eq. \ref{eq:eqCor} plotted as a difference compared to PB theory. Note that the profiles are organized by $z_\mathrm{counterion}:z_\mathrm{coion}$ in each row and by $C_\mathrm{ref}$ in each column.}
\label{fig:figS8} 
\end{figure*}

\begin{figure*}[b] 
\includegraphics[width=0.8\linewidth]{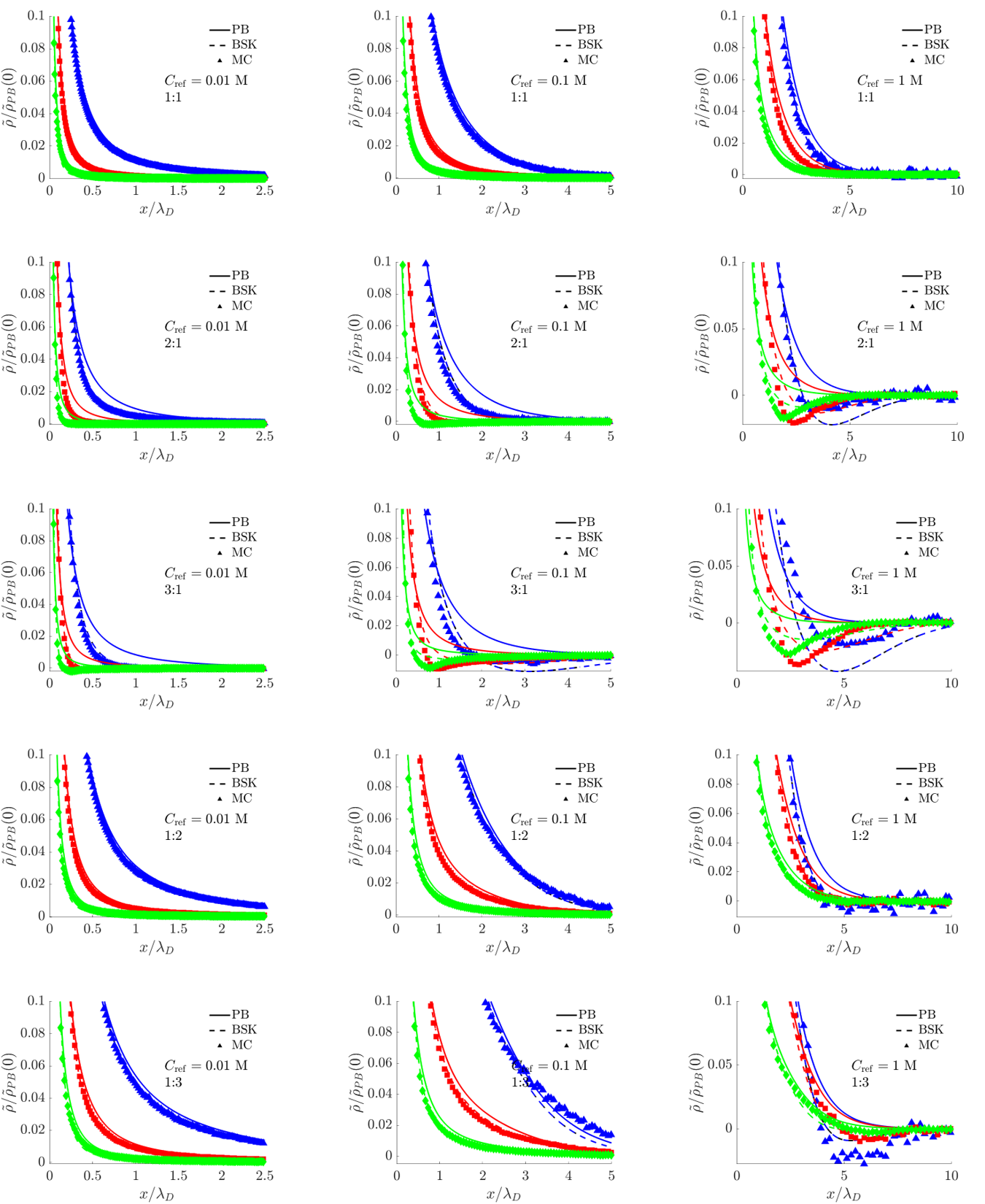}
\caption{Comprehensive comparison of BSK theory with MC simulations from \cite{valisko2018systematic} for the correlation length scaling from Eq. \ref{eq:eqCor} plotted as the total charge density. The axis are zoomed in to isolate the extent of overscreening. Note that the profiles are organized by $z_\mathrm{counterion}:z_\mathrm{coion}$ in each row and by $C_\mathrm{ref}$ in each column.}
\label{fig:figS9} 
\end{figure*}

A set of comparisons for the fitted correlation length scaling in Eq. \ref{eq:eqCor} is plotted in Figs. \ref{fig:figS8} and \ref{fig:figS9}, showing more uniform agreement than the correlation length of $\ell_c=0.50 R_\mathrm{hole}$. Even given the fitted correlation length, it is apparent that for the case where the surface charge density goes to zero, the definition of the correlation length must change. At very low surface charges, the correlation length will be governed by the bulk correlation length of charges. Furthermore, here we only assume a constant correlation length, but it is possible for this length to be dependent on the local concentration or distance from a surface. More studies are necessary to elucidate the spatial dependence of the correlation length, especially when considering multicomponent mixtures with varying ion valency. 

\bibliography{library}